2025-08-28

# Giant shot noise in superconductor/ferromagnet junctions with orbital-symmetry-controlled spin-orbit coupling

César González-Ruano[1,2,#], Chenghao Shen[3,4,#], Pablo Tuero[1,#], Coriolan Tiusan[5], Yuan Lu[6], Jong E. Han[3], Igor Žutić[3,*] & Farkhad G. Aliev[1,7,8*]

[1]Departamento Física de la Materia Condensada C-III, Universidad Autónoma de Madrid, 28049 Madrid, Spain
[2]Institute for Research and Technology (IIT), School of Engineering (ICAI), Universidad Pontificia Comillas, C/Alberto Aguilera, 23, 28015, Madrid, Spain
[3]Department of Physics, University at Buffalo, State University of New York, Buffalo, NY 14260, USA
[4] Eastern Institute for Advanced Study, Eastern Institute of Technology, Ningbo, Zhejiang 315200, China
[5]Department of Solid State Physics and Advanced Technologies, Faculty of Physics, Babes-Bolyai University, Cluj-Napoca, 400114, Romania
[6] Université de Lorraine, CNRS, Institut Jean Lamour, F-54000 Nancy, France
[7] Instituto Nicolás Cabrera (INC), Universidad Autónoma de Madrid, 28049 Madrid, Spain
[8] Condensed Matter Physics Institute (IFIMAC), Universidad Autónoma de Madrid, 28049 Madrid, Spain
[#] These authors contributed equally

By measuring the shot noise, a consequence of charge quantization, in super-conductor/insulator/ferromagnet (V/MgO/Fe) junctions, we discover a giant increase, orders of magnitude larger than expected. The origin of this giant noise is a peculiar realization of a superconducting proximity effect, where a simple superconductor influences its neighbors. Our measurements reveal largely unexplored implications of orbital-symmetry-controlled proximity effects. The importance of orbital symmetries and the accompanying spin-orbit coupling is manifested by an unexpected emergence of another superconducting region strikingly different from the parent superconductor. Unlike vanadium's common spin-singlet superconductivity, the broken inversion symmetry in V/MgO/Fe junctions and the resulting interfacial spin-orbit coupling leads to the formation of spin-triplet superconductivity across the ferromagnetic iron. Here we show that the enhanced shot noise, known from Josephson junctions with *two superconductors*, is measured even in a *single superconductor*, this discovery motivates revisiting how the spin-orbit coupling and superconducting proximity effects can transform many materials.

## INTRODUCTION

A common goal in science and technology is to suppress noise and increase the signal-to-noise ratio[1-3]. However, noise also enables elucidating subtle phenomena, hidden from other experimental probes[4-9]. A striking example is shot noise, which can be used to characterize strongly-correlated systems, strange metals, quantum entanglement, or fractionally charged quasiparticles[7-11]. Since shot noise stems from charge quantization, the fractional quantum Hall effect can be characterized by the reduced shot noise, while in superconductors, where the charge is added in Cooper pairs, even doubling of the normal-state shot noise is possible. The concept of proximity effects, where a given material is influenced by its neighbors, has been known for superconductors for over ninety years[12,13]. In Josephson junctions, where the superconductivity from the *two superconductors* can leak across the nonsuperconducting region, the proximity effects have been identified through the observation of multiple reflections of Cooper pairs and accompanied by a large, measured shot noise[14-16]. While proximity effects have been extensively studied in junctions with a *single* superconducting region, so far, there have been no measurements of the enhanced shot noise in such structures.

Conventional superconductivity is a condensation of Cooper pairs, consisting of electrons with opposite spins. Such spin-singlet superconductivity competes with ferromagnetism, which tends to align the electron spins parallel. In superconductor/ferromagnet (S/F) junctions this competition lowers the superconducting critical temperature $T_C$ and strongly suppresses the proximity effect, such that the characteristic length (~few nm) for superconductivity leaking into F is orders of magnitude shorter than in the S/normal metal (N) junctions**[17-21]**. However, if Cooper pairs are transformed to have equal (parallel) spins, the resulting equal-spin-triplet superconductivity would have the ability to coexist as a long-range triplet (LRT) with ferromagnetism, to transfer dissipationless spin currents, desirable for superconducting spintronics, and even to support elusive Majorana states for fault-tolerant topological quantum computing[17,22].

Rather than seeking the desired spin-triplet superconductivity in a single material such as $Sr_2RuO_4$, until recently expected to be a prime candidate[23], there is a growing effort to realize proximity-induced spin-triplet superconductivity in suitable junctions**[17-19,21,22]**. For two decades, such an approach to achieve LRT was focused on using complex ferromagnetic multilayers, typically relying on noncollinear or spiral magnetization[21,24-26] or half-metals[27-30]. More recently, interfacial spin-orbit coupling (SOC) was considered as an alternative platform, where proximity-induced spin-triplet superconductivity could be supported even with a single conventional ferromagnet[17,31]. In our work, using shot noise spectroscopy in high-quality epitaxial junctions, the significance of this SOC is even more striking. In the normal state the SOC determines the transport properties, while in the superconducting state we address the long-standing search for the proximity-induced Josephson effect[32], enabled by the additional SOC-supported spin-triplet superconducting region.

**RESULTS**
Our choice of epitaxial V(100)/MgO/Fe(100) junctions appears surprising. The incompatible orbital symmetries in the electronic structure of V(100) and Fe(100), suggest that their junctions are nonconducting in the low-bias regime[33-35]. Early experiments on heteroepitaxial superconducting Fe/V/Fe junctions reveal the importance of the relevant orbital symmetries and how they can determine desirable spin-dependent transport properties[36].

At the Fermi level $E_F$, $\Delta_2$ orbital symmetry of V is absent for Fe, which is characterized by $\Delta_1$ symmetry, shown in Fig. 1a. Since MgO is an insulator filtering out $\Delta_2$ and producing giant tunneling magnetoresistance (TMR) in Fe/MgO/Fe junctions[33-35,37,38], proximity-induced superconductivity from V across MgO into Fe seems even less likely. However, structural inversion asymmetry in our junctions leads to interfacial SOC which accompanies the effective $\Delta_2$ barrier in crystalline MgO due to its filtering effect. The SOC-induced spin-flip scattering mixes $\Delta_2$ and $\Delta_1$ symmetries and creates a mechanism for electron tunneling above $T_C$ across the junction at low bias, shown in Fig. 1a. Our first-principles calculations (see Supplementary Information, Section I, SI-I) also confirm the presence of Rashba SOC at the V/MgO interfaces. Another low-bias contribution in these epitaxial junctions with conserved wave vector parallel to the interfaces, $k_\parallel$, comes from "hot spots" for normal incidence at $k_\parallel = 0$ (the $\Gamma$ point) which provides high transmission through MgO[33]. A simple picture for the normal-state transport in this junction is described by an equivalent resistance, $R_{eq} = R_{SOC} + R_{MgO}$, in which the resistance of the symmetry-related SOC barrier, $R_{SOC}$, is much larger than the resistance from the conventional barrier strength of the MgO region, $R_{MgO}$.

This picture of the tunneling dominated by symmetry-enforced spin filtering, rather than by the barrier strength, is confirmed by the measured low-bias differential conductance $G$ shown in **Fig**. 1b, revealing the key role of SOC at the V/MgO interface. Compared to junctions without such an interface, $G$ is reduced by two (three) orders of magnitude with one (two) interface(s). For in-plane magnetization in a V/MgO/Fe/MgO/Fe/Co junction, shown in Fig. 1c, the obtained $TMR = (G_P - G_{AP})/G_{AP} \sim 40\%$, where $G_P$ ($G_{AP}$) corresponds to parallel (antiparallel) magnetization in the two Fe regions, signals highly spin-polarized electrons. By excluding the nonmagnetic V/MgO region, we have shown an even higher TMR~330% in Fe/MgO/Fe/Co junctions[38]. Using shot noise measurements, we can also exclude the role of pinholes in determining $G$ (details of sample fabrication, measurements, and characterization can be found in Methods, **SI-II** and Ref. 31). To further support our conductance measurements across different junctions, it is helpful to consider a schematic illustration of the role of dominant orbital symmetries and their SOC-induced mixing in Figs. 1d-1f. The highest measured conductance in Fig. 1b is found for the Fe/MgO/Fe-based junction in which the dominant $\Delta_1$ symmetry is shared by all the regions and the related transport does not experience the symmetry mismatch (no SOC barrier). However, for junctions with regions characterized by other orbital symmetries, the spin-filtering exclusion from the MgO can be overcome by SOC mixing and accompanied by a large $R_{SOC}$, consistent with the results in Fig. 1b.

In the superconducting state, as shown in Fig. 2a, transport is distinguished by Andreev reflection, providing the microscopic mechanism for proximity-induced superconductivity[13,17]. In conventional Andreev reflection an electron is reflected backwards and converted into a hole of opposite charge and spin. With no interfacial barrier, this implies the doubling of the normal-state $G_N$: two electrons are transferred across the interface into S, where they form a spin-singlet Cooper pair[39]. Because of spin polarization $P$ in F, not all electrons can find a partner of opposite spin to undergo Andreev reflection[17,40]. Together with the normal (ordinary) reflection at the interfacial barrier, such a finite $P$ suppresses the Andreev reflection and reduces $G$[40] at applied bias $V < \Delta/e$ below the effective superconducting gap $\Delta \sim 1.05$ meV, where -e is the electron charge. A small peak in $G(V \sim \Delta/e)$ and a substantial value of $G(V=0)$ (Fig. 2a) suggest that S/F is not a typical tunnel junction and has only a moderate interfacial barrier strength. The inset in Fig. 2a indicates $T_C \sim 4$ K, as identified by the measured temperature-dependent conductance.

A high-quality MgO barrier defines the location of the voltage drop and thus enables accurate shot noise measurements, previously absent in S/F junctions. Unlike the measured $G(V)$, similar to what was observed in various superconducting structures, the shot noise in Fig. 2b for the same S/F junction shows an unprecedented giant low-bias increase at $T = 0.3$ K $< T_C$, orders of magnitude larger than theoretically expected. With fluctuations due to the discreteness of the electrical charge, it is common to introduce the current shot noise power $S_{I,max} = 2q<I>$ transferred in discrete units of charge q, where $<I>$ is the average current[10,11,41,42]. To describe the ratio between the shot noise and the conductance, it is useful to introduce the Fano factor[10,11] $F = S_I/(2eG|V|)$. This Fano factor also gives the effective charge responsible for the shot noise[10,11]. For a fully random (Poisson) process of uncorrelated electrons, $S_{Poisson} = 2e<I>$, $F$ in the normal state attains at maximum $F = 1$, while $F = 2$ for superconducting tunnel junctions[11] signals that $|q| = 2e$ since the shot noise originates from the transfer of Cooper pairs. With a finite circuit resistance, our measurements also include voltage fluctuations with the resulting voltage shot noise power[10,42,43] $S_V = F2e<I>/G^2$ (SI-II provides the expression for $T > 0$), shown together with the expected maximum normal-state value for $F = 1$. Remarkably, at $T = 0.3$ K and $eV < \Delta$, we can infer $F > 100$, as if a giant effective charge $|q| > 100$ e is responsible for the observed shot noise! This striking behavior is the hallmark of the superconducting state, while at $eV > \Delta$ or $T > T_C$, $S_V$ approaches the Poisson value for $F = 1$ (green line). Another distinguishing feature of the giant $S_V$ in the superconducting state is its independence of frequency $f$, shown in Fig. 2c over two orders of magnitude in the frequency range. In contrast, the contribution of vortices[43] is reflected in the $1/f$-dependent part of the noise, decreasing with $f$ and increasing with $T$, when $T$ approaches $T_C$ (SI-III).

It is helpful to compare our prior conductance and shot noise results for S/I/F (V/MgO/Fe) junction with the measurements on a control S/I/N junction (V/MgO/Au). As expected, in Fig. 2d we find that the low-bias conductance is suppressed less in V/MgO/Au than that in V/MgO/Fe junction, consistent with the nonmagnetic Au and $P = 0$, as the Andreev reflection does not experience a large suppression from $P = 0.7$ at the Fe/MgO interface. Both junctions are not in a typical tunneling regime and they share only a small peak in $G$, known to appear at $V \sim \Delta/e$. However, the separation of the two peaks is larger in V/MgO/Fe where, considering the commonly expected competition between ferromagnetism and superconductivity[17,21], such a separation and the related superconducting gap should be reduced compared to the one in nonmagnetic V/MgO/Au. Since in both junctions we see that the superconducting gap exceeds the values expected for V itself, a larger peak separation in V/MgO/Fe would be consistent with a slightly larger proximity-induced gap than in V/MgO/Au (see discussion below). While the changes in the two corresponding gaps are moderate, turning to the comparison of the shot noise in Fig. 2e, we see a drastic increase in the measured low-bias shot noise of V/MgO/Fe junction. These trends in the conductance and shot noise by replacing F by N region can be partially understood by recognizing the importance of the epitaxial growth and high-quality interfaces in V/MgO/Fe, while V/MgO/Au is an example of nonepitaxial growth leading to the highly textured interface and suppressed filtering due to different orbital symmetries, as depicted in Fig. 2f and SI-II. That the apparent superconducting gap is not more suppressed in V/MgO/Fe could signal the presence of spin-triplet superconductivity coexisting with ferromagnetism[17,31].

The correlations among electrons, due to Coulomb repulsion and the Pauli exclusion principle, reduce the shot noise below its Poisson limit $F = 1$ ($F = 2$) in the normal (superconducting) state[10,11]. For example, in the fractional quantum Hall effect, the measured $F = 1/3$ signals the characteristic fractional charge[4]. Rare exceptions where the shot noise is enhanced compared to the Poisson value[14-16] are typically observed in junctions with multiple superconducting regions without ferromagnets. To examine such excess shot noise, we consider a V/MgO/Fe/MgO/Fe/Co junction, used also to measure TMR in Fig. 1c. With the two different F regions we can control the orientation of the Fe magnetization while the higher-coercivity Fe/Co magnetization remains fixed and serves as a sensor of the Fe magnetization through the measured *TMR* signal.

For the V/MgO/Fe/MgO/Fe/Co junction in Fig. 3a we see another example of a giant increase in the subgap shot noise, strongly suppressed by the out-of-plane (OOP) applied magnetic field, $H_{OOP}$. This is further examined in Fig. 3b, which shows the suppression of the maximum value of the subgap shot noise with both $H_{OOP}$ and in-plane (IP) $H_{IP}$. A stronger $S_V$ suppression with $H_{OOP}$ than with $H_{IP}$ is expected, just as for the OOP $H$-suppression of $\Delta$ (in the inset). However, the in-plane anisotropy (between [100] and [110] orientations for both $S_V$ and $\Delta$) is surprising for conventional spin-singlet superconductivity and could instead signal an induced spin-triplet superconductivity. We also observe a slight increase in the noise with $H_{IP}$, which can be attributed to the field-suppression of inherent magnetic textures. Similar trends in the suppression with $V$ and $H$ are also reproduced in the measured Fano factor in Fig. 3c, which reaches nearly $F = 200$.

To seek a possible explanation of this peculiar behavior, we recall that normal-state transport in V/MgO/Fe-based junctions and our first-principles calculations confirm the presence of the interfacial SOC. Given the measured $H_{IP}$ noise anisotropy, such SOC influences the superconducting state and leads to spin-flip Andreev reflection with equal spin for electrons and holes, which is responsible for spin-triplet Cooper pairs[17,44,45]. As a result, SOC can partially counteract the suppression of Andreev reflection due to the high spin polarization, $P \sim 0.7$-$0.8$, of the Fe/MgO contact[31] and the normalized interfacial barrier strength $Z$, such that the resulting low-bias $G$ is higher than without SOC[45]. This is consistent with the data from Fig. 2a, which indicate only a moderate $Z \sim 1$ ($Z = 0$ for perfect transparency and $Z >> 1$ for a tunnel junction). However, when we theoretically explore the shot noise in S/F junctions over a large parameter space for normalized barrier and Rashba SOC strength $Z$ and $\lambda$ (see Methods and SI-IVA, B) respectively, the maximum value is $F = 2$ (SI-IVC), orders of magnitude smaller than in Figs. 2b or 3c. With our analysis of the calculated shot noise in an effective 1D system, in the absence of spin polarization, it is possible to analytically express different scattering coefficients for the S/N junction and use them to evaluate the enhanced Fano factor for S/N/S Josephson junction (SI-IVD).

The interplay of ferromagnetism and SOC is expected to generate proximity-induced spin-triplet superconductivity, but that alone cannot explain a huge excess of shot noise. Instead, prior examples of large shot noise[14-16] suggest a resonant behavior which could naturally occur in Josephson junctions, with two superconducting regions and the formation of Andreev bound states, defined by multiple Andreev reflections (MAR)[20,46]. In the simple case of two identical superconducting gaps and $Z = 0$, one expects n = $\Delta/eV$ Andreev reflections, as if the charge transfer and the underlying shot noise could be described by a composite object with effective charge $|q| =$ ne[47]. Support that only a single S region in nonmagnetic junctions could display properties of proximity-induced Josephson effect[32,47,48] motivates us to revisit the understanding of S/F junctions. Proximity-induced spin-singlet superconductivity is strongly suppressed in F such as Fe. However, its SOC-induced spin-triplet counterpart could coexist with ferromagnets, and we also consider that it is accompanied by an effective superconducting gap $\Delta_F$, and therefore supports MAR and enhances $F$ beyond 2. The equal-spin superconducting correlations derived from the superconducting condensate inside V evolve dynamically through the interplay between SOC, symmetry-dependent tunneling, and the exchange interaction. The resulting phase evolution effectively decouples the proximity-induced superconducting correlations from the parent superconductor and supports MAR, which is also observed from the preformed pairs[6,7]. This is unlike the usual proximity effects in S/F (S/N) junctions, where the proximitized order parameter is directly phase locked to that of S[21].

A sketch of the physical mechanism for the giant shot noise and induced $\Delta_F$ is depicted in Fig. 4a. The considered multiple electron-hole reflections are consistent with the abundance of these quasiparticles with $\Delta_1$ symmetry at the Fermi level of Fe(001), which easily tunnel through the MgO due to hot spots in momentum space. Multiple states with complex wave vectors in the MgO lead to interference effects in $G$[31]. The Fano factor gives the effective charge transfer through MAR across the apparent proximity-induced Josephson junction, formed by the vanadium *s*-wave superconductor and the resulting spin-triplet correlations in Fe. With this picture, we generalize the MAR[15] calculation for $G$ and the Fano factor to include the influence of SOC and normal reflections.

This simple phenomenological model, nevertheless, captures several important experimental observations. With the calculated $G$ in Fig. 4b we see that SOC enhances the interfacial transparency, consistent with the measurements from Fig. 2a which indicate only a moderate $Z \sim 1$, rather than the conventional tunnel junction with $Z \gg 1$. By explicitly including $\Delta_F > 0$, we find that MAR-modified $G$ no longer has the usual peak at the pure superconducting gap for vanadium, $\Delta_S$, but is shifted to higher values, $eV_{peak} = \Delta_S + \Delta_F$. Considering the broadening effect on the $G$ – $V$ curve due to thermal smearing and inelastic scattering in experiments, this peak position could be further shifted to higher energies[49-51], leading to superconducting gaps ($\Delta_{Fe}$, $\Delta_{Au}$) in Fig. 2d that exceed the expected BCS value of $\Delta_S = 0.7$ meV for vanadium with a measured critical temperature of $T_C \sim 4$ K. Assuming that the spin-triplet gap is negligible in the S/I/N (V/MgO/Au) control junction and, based on the measured gaps ($\Delta_{Fe}$, $\Delta_{Au}$) in Fig. 2d, we find $\Delta_{Fe} / \Delta_{Au} = (\Delta_F + \Delta_S) / \Delta_S \approx 1.2$, from which we estimate the induced gap $\Delta_F \approx 20\% \Delta_S$. With these parameters, we achieve an excellent fit to the experimental conductance (Fig. 2a, SI-IVF). This peak shift effect provides additional evidence for the proximity-induced triplet gap and the Josephson junction-like behavior in the S/I/F (V/MgO/Fe) junction.

Considering next the Fano factor, which in Fig. 4c is represented by |q|, the calculated effective charge transfer (SI-IVE), we confirm the essential role of proximity-induced $\Delta_F$, while a two-fold change of Z has only a very small influence. A finite $\Delta_F$, through MAR, allows for a large number of electron pairs to be transferred into the superconducting lead, resulting in a giant $F$ near vanishing $V$. In the opposite large-bias limit, $eV > \Delta$, |q| approaches the uncorrelated limit e, as the current is carried by independent quasiparticles. However, with $\Delta_F = 0$, the higher-order Andreev reflections alternate between the electron and hole pairs transferred into the superconducting lead, without any giant Fano factor. Even at vanishing $V$, the calculated |q| retains its conventional value of 2, known for S/N junctions[10]. The inset of Fig. 4c shows that the magnitude of the low-bias Fano factor grows with the increasing proximity-induced $\Delta_F$, together with an increased bias value for the peak position in $G$. Even in the extreme limit of $\Delta_F = \Delta_S$, our calculated results underestimate the measured Fano factor. We attribute this limitation to our simple and transparent description, which neglects the random scattering from the SOC barrier and the destructive interference[52] suppressing the current (and therefore increasing $F$). Including these omitted effects could provide closer agreement with the measured Fano factor. Nevertheless, our theoretical framework already addresses the observed major puzzle. We provide a mechanism to exceed the

expected theoretical limit $F = 2$ (shown in gray) with a single S region, while using the same parameters that describe the measured $G(V)$ from Fig. 2a.

The significance of SOC and orbital symmetry selection in high-quality epitaxial junctions is further verified from our control V/MgO/V measurements (SI-II). One may expect that this conventional S/I/S Josephson junction would support an even larger Fano factor than in S/I/F junctions that we have studied. Indeed, experiments in NbN/MgO/NbN junctions confirm MAR and an enhanced shot noise[15]. Instead, with different orbital symmetries in V/MgO/V junctions, $\Delta_2$ in vanadium and $\Delta_1$ in MgO, both the conductance and subgap shot noise are suppressed several orders of magnitude as compared to V/MgO/Fe junctions.

These results suggest several important future opportunities. The proximity-induced Josephson effect in N/I/S junctions[47,48] was observed through $I$-$V$ curves or a zero-bias conductance peak (ZBCP), reproducing properties of conventional S/I/S Josephson junctions. Since the origin of similar ZBCP observations[53] continues to be studied and attributed to resonant effects[54,55], even without the proximity-induced second superconducting region, our noise spectroscopy could distinguish various scenarios and detect the proximity-induced Josephson effect. Our focus on the simple and transparent theoretical approach invites future theoretical extensions. Resonant effects alone would not support MAR and preclude our observed giant shot noise. Another ZBCP implication is its signature of Majorana states[55] in spin-triplet topological superconductivity considered for fault-tolerant quantum computing[17,22]. However, an extrinsic ZBCP origin remains debated[22,57] and shot noise spectroscopy could help to identify Majorana states[58].

While Fe/MgO-based junctions have been extensively studied[34,35], from commercial applications[59] to integrating spintronics, electronics, and photonics[60], there is only a limited understanding about their all-epitaxial growth with superconductors[17,31]. Since we show that even in the normal-state for these junctions their resistance is dominated by the symmetry-related SOC barrier (recall Fig. **1b**), rather than the conventional barrier associated with the MgO regions, this motivates further studies to explore the superconducting spintronics in all-epitaxial superconductor junctions with ferromagnets where the spin-triplet proximity and spin currents could be controlled by SOC. The role of SOC in proximity-induced spin-triplet topological superconductivity is well studied using semiconductor nanostructures[17,22], but often overlooked in ferromagnetic junctions[21,28-30]. Shot noise spectroscopy could overcome these uncertainties, elucidating the role of SOC in systems where there remains a debate about the induced long-range spin-triplet superconductivity[30,61,62]. Our findings pertain also to materials design and emergent phenomena through various proximity effects where buried interfaces play a crucial role[13]. While probing such interfacial properties poses a challenge for many scanning probes, our work demonstrates that even buried interfaces are directly accessible to noise spectroscopy.

**Data availability**

The data supporting the findings of this study are available within the Article and its Supplementary Information files. This includes main trends in the theoretical results, given analytically in the Supplementary Information. The data for the calculations shown in figure 4 are available at the Figshare repository, under the following URL: doi.org/10.6084/m9.figshare.29988889. The raw data files for the experimental results are available from the corresponding authors upon request. Source data are provided with this paper.

**Code availability**

The computational code required to reproduce the theoretical calculations is available from the corresponding author upon reasonable request.

## METHODS

**Sample preparation and characterization**

The magnetic tunnel junction (MTJ) multilayer stacks were grown by molecular beam epitaxy (MBE) in a chamber with a base pressure of $5 \cdot 10^{-11}$ mbar following the procedure described in Ref.[63]. The samples were grown on (100) MgO substrates with a 10 nm thick seed of anti-diffusion MgO underlayer on the substrate to trap the C from it before the deposition of the Fe or V electrodes. Then the MgO insulating layer was epitaxially grown by e-beam evaporation with the thickness of approximately 2 nm and followed by the rest of the layers. Each layer was annealed at 450 ºC for 20 mins for flattening. After the MBE growth, all the MTJ multilayer stacks were patterned in square junctions of a size from 10 to 40 μm (with the diagonal along [100]) by UV lithography and Ar ion etching, controlled step-by-step *in situ* by Auger spectroscopy. For the considered samples, the measured properties did not depend on the lateral size of our junctions.

For S/F junctions studied in the main text, the layer structure was V(40 nm)/MgO(2 nm)/Fe(10 nm). For S/F/F junctions, the base part V/MgO/Fe was the same, with an extra MgO(2 nm) barrier and a Fe(10 nm)/Co(20 nm) hard ferromagnetic layer grown on the top. The F/S/F junction structure was Fe(45 nm)/V(40 nm)/MgO(2 nm)/Fe(10 nm)/Co(20 nm), and the control S/S junctions had two 40 nm vanadium layers separated by a 2 nm thick MgO barrier.

**Experimental measurement methods**

The measurements were performed inside a JANIS® $^3$He cryostat with a base temperature of 0.3 K. The magnetic field was varied using a 3D vector magnet consisting of one solenoid (Z axis) with $H_{max}$ = 3.5 T and two Helmholtz coils (X and Y axis) with $H_{max}$ = 1 T. In our system, the different magnetic states can be distinguished by looking at the resistance of these S/F/F junctions, so the relative orientation between the two F electrodes can be measured. The magnetoresistance measurements were performed by first setting the magnetic field to the desired value, then applying a positive and negative current up to the desired voltage (5 mV unless otherwise stated), and averaging the absolute values of the measured voltage for the positive and negative current, obtaining a mean voltage which was used to calculate the resistance at that point.

The electron current was supplied by a Keithley 220 low-noise current source. The voltage signal (DC + fluctuations) produced through the samples by this current was duplicated and then amplified, first by a homemade preamplifier (based on two INA111 instrumentation amplifiers), and then both channels were amplified again with two Stanford Research SR560 commercial amplifiers with the same configuration, which also filter out the DC component of the signal and apply a bandpass filter with a range that can be varied from 0 Hz to 1 MHz. For the differential conductance/resistance, the voltage was measured from one of the channels (which can be switched) with a Digital Multimeter PCI Board (DMM-552-PCI). Finally, both signals were sent to a Stanford Research SR785 spectrum analyzer which has a bandwidth of 102.4 kHz. The two inputs contain the sample noise signal $\delta V_S$ and an extrinsic noise contribution $\delta V_{amp}$ coming from the wires and the amplifiers. The noise from the sample is the same in both channels, but the extrinsic component due to the electronics of the voltage is uncorrelated between each channel, although of a similar magnitude. All the unnecessary electronic equipment was turned off to reduce external noise sources. The current was varied step by step and the voltage signal was duplicated and analyzed in the SRS785 performing cross-correlation to exclude the extrinsic part of the signal[42,64,65]. The cross-correlation technique allows us to decrease the background noise base level and to evaluate the frequency dependence of the noise power (see **Fig. 1c** in the main text) to exclude possible random telegraph noise contributions. To attain the same objective, other shot noise measurement systems with the single channel amplification only measure the noise within a narrow frequency band at higher frequencies[42].

**Theoretical methods**

To compute the transport noise, we extend a widely used approach for obtaining *I-V* characteristics and $G$[39,46] to decompose the electron distribution according to the number of Andreev reflections[15] in the Josephson current by keeping track of the net electron charge in each transmitted electron wavefunction. The effective charge in multiple Andreev reflections comes from coherent pair transport that greatly exceeds the charge of a pair of electrons. We first obtain scattering coefficients for the normal and Andreev reflection as well as the transmission in a single S/F junction with the SOC[44,45] from the microscopic Bogoliubov-de Gennes Hamiltonian as a $4 \times 4$ matrix

$$\widehat{H} = \begin{pmatrix} \widehat{H}_e & \Delta_S \Theta(z) I_{2x2} \\ \Delta_S^* \Theta(z) I_{2x2} & \widehat{H}_h \end{pmatrix}, \quad (1)$$

$$\text{with } \widehat{H}_e = -\frac{\hbar^2}{2m^*}\nabla^2 - \mu - \frac{\Delta_{xc}}{2}\widehat{\mathbf{m}} \cdot \boldsymbol{\sigma}\,\Theta(-z) + \left[V_0 d + \alpha(k_y \sigma_x - k_x \sigma_y)\right]\delta(z) \quad (2)$$

and $\widehat{H}_h = -\sigma_y \widehat{H}^*_e \sigma_y$, where $\widehat{H}_e, \widehat{H}_h$ are the single-particle Hamiltonians for electrons and holes, while $I_{2x2}$ is the two-dimensional unit matrix. $\Delta_S$ is the spin-singlet superconducting gap for vanadium, $\Theta(z)$ is the step function in the $z$ direction, perpendicular to the S/F interface at $z = 0$, $m^*$ is the effective electron mass, $\mu$ the chemical potential, $\Delta_{xc}$ the exchange spin splitting along the direction $\hat{\mathbf{m}}$ of the magnetization. The interfacial scattering is modeled by a delta-function potential scattering, characterized by the effective barrier height $V_0$ and thickness $d$ and the Rashba SOC strength $\alpha$. $\sigma_{i=x,y,z}$ is the $2 \times 2$ Pauli matrix for the electron spin. It is convenient to introduce the dimensionless parameters for potential barrier and SOC strength[44, 45]

$$Z = \frac{V_0 d m^*}{\hbar^2 k_F}, \qquad \lambda = \frac{2\alpha m^*}{\hbar^2}. \tag{3}$$

In our calculations, we use $m^* = m_0$, where $m_0$ is the free-electron mass, and the chemical potential $\mu = 2.5$ eV. The spin polarization in the ferromagnet is $P = \Delta_{xc}/(2\mu) = 0.7$.

The scattering coefficients from the single junction are then used to compute the conductance and the effective charge $|q|$ in the F/S/F Josephson junction under bias by extending the approach[39, 46], which neglects SOC and spin polarization. This formulation generalizes the electron distribution function[39] $f(E)$ at energy $E$ decomposed as

$$f(E) = \sum_{N=0}^{\infty} \sum_{m=-\infty}^{\infty} g_m^N(E), \tag{4}$$

where the function $g_m^N(E)$ is resolved after $N$ reflections from either of the S/F interfaces and $m$ is the net electric current transmitted in the leads. Recursion relations for $g_m^N(E)$ after an additional reflection with the scattering coefficients from the S/F interfaces are derived in **SI-IVA,B**. The effective charge in the current noise is then computed with the transmission coefficient $T(E)$ as

$$\langle |q| \rangle = \frac{\sum_m |m| I_m}{\sum_m I_m} \text{ with } I_m = \sum_{N=0}^{\infty} \int T(E) g_m^N(E) dE. \tag{5}$$

A detailed formulation and more rigorous expressions are presented in **SI-IVA,B**, with an example demonstrated in a simple 1D limit in **SI-IVD**.

Requests for materials should be addressed to Farkhad Aliev or Igor Žutić.

**Acknowledgements** The work in Madrid was supported by Spanish Ministry of Science and Innovation (PID2021-124585NB-C32, TED2021-130196B-C22, PID2024-155399NB-I00) (F.G.A.) and Consejería de Educación e Investigación de la Comunidad de Madrid (Mag4TIC-CM Ref. TEC-2024/TEC-38) (F.G.A.). grants. Financial support from the Spanish Ministry of Science and Innovation, through the María de Maeztu Programme for Units of Excellence in R&D CEX2023-001316-M is also acknowledged (F.G.A.). The work in Buffalo was supported by the US Department of Energy (DOE) Office of Science Basic Energy Sciences (BES) award no. DE-SC0004890 (C.S., I.Ž.). Computational resources were provided by the UB Center for Computational Research. C. T. acknowledges the following funding sources: the project "MODESKY" ID PN-III-P4-ID-PCE-2020-0230-P, grant No. UEFISCDI: PCE 245/02.11.2021 and the BBU grants AGC30199/17.01.2025, AGC30200/17.01.2025 and GS-UBB-FF - Coriolan Tiusan/2025 (C.T.). The support by the French National Research Agency (ANR) SOTspinLED project (no. ANR-22-CE24-0006-01), CHIFTS project (No. ANR-23-CE09-0006-003)and PEPR-SPIN OptoSpinCom project (No. ANR-24-EXSP-0010) is also acknowledged (Y.L.). The growth of samples was performed using equipment from the CC-DAUM platform funded by FEDER (EU), ANR, the Region Lorraine and the metropole of Grand Nancy.

**Author contributions** Y.L. and C.T. fabricated the samples and performed structural characterization. C.G.-R. and P.T. measured and analyzed the conductance and noise with the guidance of F.G.A. C.S., J.E.H, and I.Ž. contributed to theoretical understanding and modelling. C.T. carried out ab-initio simulations. F.G.A. designed the overall experiment. F.G.A. and I.Ž. wrote the manuscript with the input and help of C.G-R., J.E.H., C.S., and C.T. All authors discussed the results and commented on the manuscript. C. G-R, P. T. and C. S. contributed equally to the manuscript.

**Competing interests** Authors declare no competing interests.

**Correspondence and requests for materials** should be addressed to Farkhad Aliev or Igor Žutić.
Corresponding authors emails and ORCID:
farkhad.aliev@uam.es https://orcid.org/0000-0002-1682-3306
zigor@buffalo.edu https://orcid.org/0000-0003-2485-226X

Rest of authors ORCID:
César González-Ruano https://orcid.org/0000-0002-3534-823X
Chenghao Shen https://orcid.org/0000-0002-8545-6220
Pablo Tuero https://orcid.org/0009-0000-7022-5326
Coriolan Tiusan https://orcid.org/0000-0003-1338-3867
Yuan Lu https://orcid.org/0000-0003-3337-8205
Jong E. Han https://orcid.org/0000-0002-5518-2986

**Main Figures**

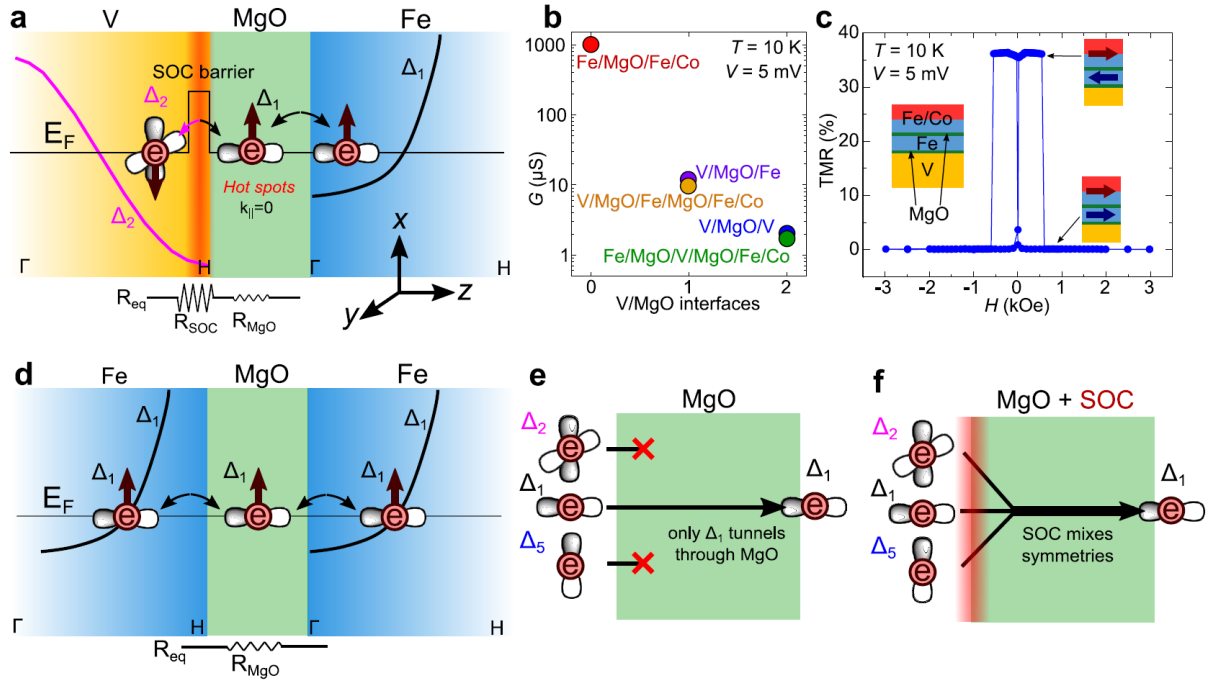

**Fig. 1| Orbital symmetry change and normal-state transport in epitaxial V/MgO/Fe-based junctions. a,** The main conductance bands, labeled with their respective orbital symmetries, are superimposed to each region. Arrows denote electron spin. At the Fermi level, $E_F$, in vanadium, only electrons with $\Delta_2$ symmetry are present, while they are absent in iron. Therefore, a symmetry change is necessary for the electron transport across the V/MgO/Fe junction. This is enabled by the Rashba spin-orbit coupling (SOC) at the V/MgO interface. MgO acts both as (i) symmetry filter at $E_F$, relatively transparent for $\Delta_1$ electrons in iron at the normal incidence (vanishing wave vector along the interface, $\mathbf{k}_{\parallel}=0$), while having a strong barrier for $\Delta_2$ electrons and (ii) enabling the symmetry and spin changes allowing electron tunneling into the iron. An equivalent resistor model indicates that the SOC barrier dominates over the usual barrier from the MgO region. **b,** Typical normal-state conductance of different tunnel junctions of a lateral size 20 x 20 µm², as a function of their number of V/MgO barriers. Each dot: sample-averaged conductance. Each extra V/MgO barrier diminishes the conductance by an order of magnitude. **c,** In-plane tunnel magnetoresistance (TMR) of a spin-valve junction (inset), showing parallel and antiparallel magnetization configurations, changing with an applied magnetic field, $H$, and the typical coercive field of the hard Fe/Co magnetic layer. **d,** Across the less-resistive Fe/MgO/Fe junction, the transport is dominated by $\Delta_1$ electrons without SOC barrier. **e, f,** The absence (presence) of SOC removes (enables) orbital symmetry mixing, explaining the measured relative magnitudes of conductance in **b**.

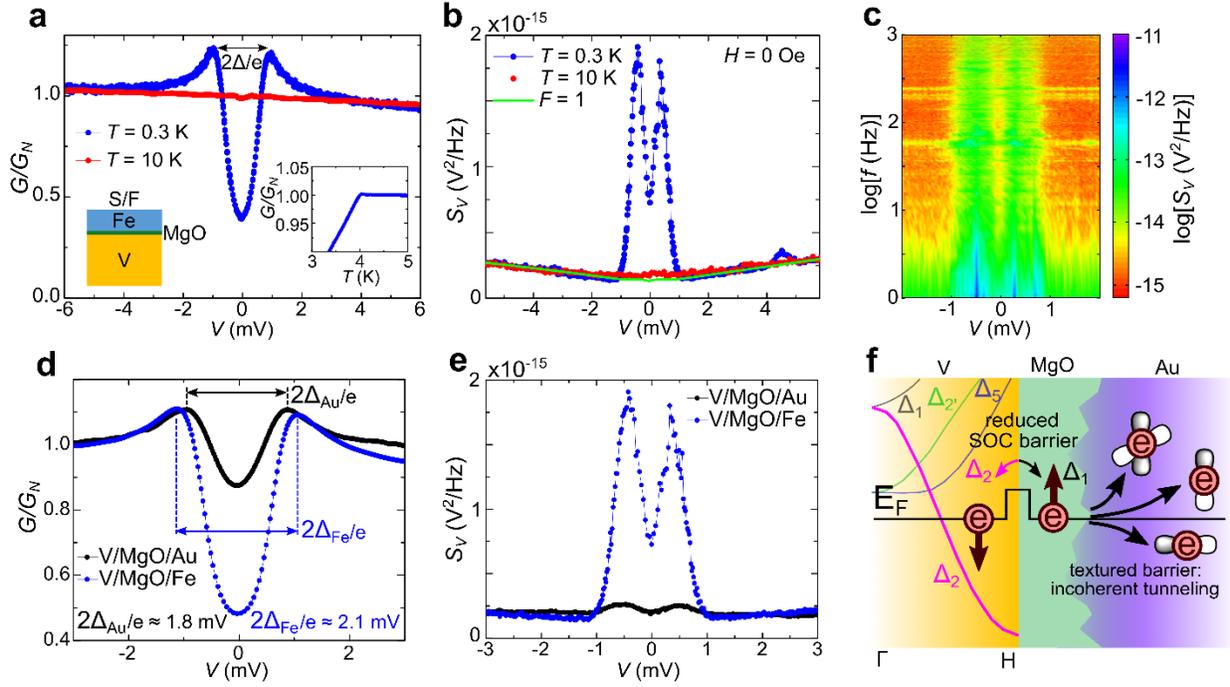

**Fig. 2| Bias dependence of the conductance and giant shot noise in a V/MgO/Fe junction. a**, Conductance for a superconductor/ferromagnet (S/F) sample (sketched in the left), above (red) and below (blue) the critical temperature of vanadium, $T_C$, normalized by the conductance at bias $V$ = -5 mV, above the effective superconducting gap, $\Delta$, -e is the electron charge. Inset: $T_C$ identified by the measured temperature-dependent subgap conductance. **b**, For the same sample, there is giant shot noise below $T_C$ (blue) at low bias, $eV < \Delta$, compared to its value above $T_C$ (red), and its corresponding theoretical maximum value given by the Fano factor, $F = 1$ (green), for the normal state. **c,** The evolution of the shot noise power with frequency, $f$, and $V$, shown in the logarithmic scale. The noise spectrum is largely $f$ independent, except near the lowest $f$. **d, e**, Comparison of the bias-dependent conductance and shot noise with the control V/MgO/Au junction, with the same normalization as in panel **a**. **f**, Schematic role of the MgO/Au nonepitaxial growth leading to the highly textured interface and suppressed filtering due to different orbital symmetries.

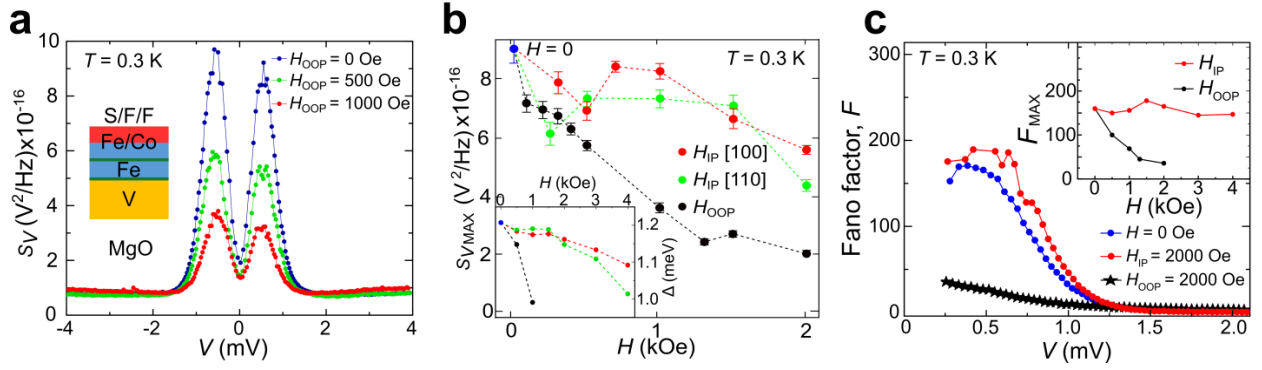

**Fig. 3| Anisotropic suppression of the giant shot noise by an applied magnetic field and bias.** The measured spin-valve junction V/MgO/Fe/MgO/Fe/Co is the same as in Fig. 1c. **a**, Out-of-plane (OOP) magnetic field effectively suppress the superconductivity and the observed low-bias shot noise. **b**, Magnetic-field dependence of the maximum noise shot shows anisotropy, not only between in-plane (IP) and OOP (black), but also for IP easy (red) and hard (green) axis, which is unexpected for a spin-singlet superconductivity. Inset: anisotropic suppression of $\Delta$ for the same directions of applied magnetic field, $H$ (same legend). **c**, Evolution of the Fano factor, $F$, with applied bias and magnetic fields. The inset: the maximum Fano factor, $F_{MAX}$, shown for a wide range of IP (red) and OOP (black) applied fields.

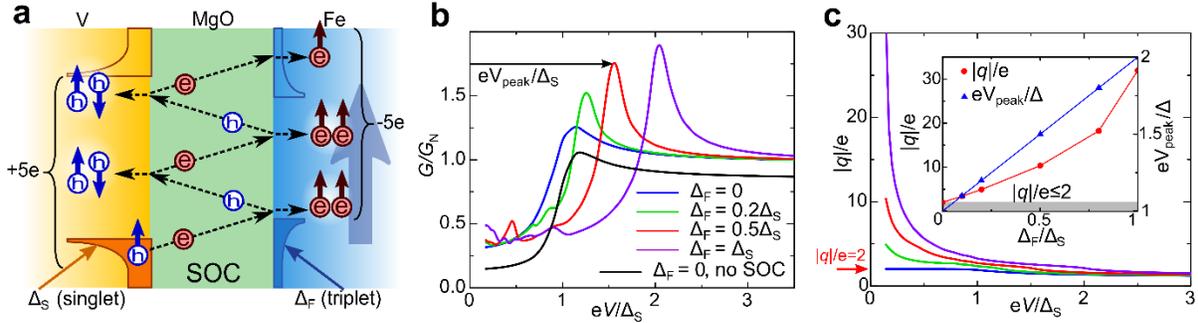

**Fig. 4| Physical mechanism with calculated conductance and shot noise. a**, Proximity-induced superconducting spin-triplet gap in Fe layer (blue), $\Delta_F$, and the interfacial spin-orbit coupling (SOC) support multiple spin-flip Andreev reflections in the V/MgO/Fe junction, which give rise to the excess charge transport (5e in the sketch) and the resulting low-bias excess Fano factor, h denotes holes. **b,** Bias-dependent conductance without (black) and with SOC (colored lines) for different relative magnitudes of the induced spin-triplet gap, $\Delta_F$, and the spin-singlet gap, $\Delta_S$, in vanadium. Without SOC, there is a reduced junction transparency and conductance for all $V$. For comparison with experimental measurements, each curve with SOC is normalized by its conductance value well above the superconducting gap. The barrier and SOC strengths are parameterized by $Z = 1$ and $\lambda = 1.2$ (see Methods). **c,** Calculated Fano factor or, equivalently, the effective charge ratio $|q|/e$, as a function of applied bias for the same parameters and a color code as given in **b**. Inset: The corresponding evolution of the effective charge ratio (at $eV/\Delta_s = 0.1$) and the conductance peak position with relative increase in $\Delta_F$. The gray area denotes the commonly expected Fano factor limited by 2.

# Supplementary Information: Giant shot noise in superconductor/ferromagnet junctions with orbital-symmetry-controlled spin-orbit coupling.

César González-Ruano,[1,2] Chenghao Shen,[3] Pablo Tuero,[1] Coriolan Tiusan,[4] Yuan Lu,[5] Jong E. Han,[3] Igor Žutić,[3] and Farkhad G. Aliev[6]

[1]*Condensed Matter Physics Department, Universidad Autónoma de Madrid, Madrid 28049, Spain*
[2]*Institute for Research and Technology (IIT) of the School of Engineering (ICAI) of Universidad Pontificia Comillas, C/Alberto Aguilera, 23, 28015, Madrid, Spain*
[3]*Department of Physics, University at Buffalo, State University of New York, Buffalo, NY 14260, USA*
[4]*Department of Solid State Physics and Advanced Technologies, Faculty of Physics, Babes-Bolyai University, Cluj-Napoca, 400114 Romania*
[5]*Institut Jean Lamour, Nancy Universitè, 54506 Vandoeuvre-les-Nancy Cedex, France*
[6]*Condensed Matter Physics Department, Instituto Nicolas Cabrera (INC) and Condensed Matter Physics Institute (IFIMAC), Universidad Autónoma de Madrid, Madrid 28049, Spain*

## I. FIRST-PRINCIPLES CALCULATION OF SPIN-ORBIT COUPLING

Our first-principles calculations were made using the WIEN2k software[1], within a Full-Potential-Linear-Augmented-Plane-Wave method. A supercell model was used to describe the V/MgO interface. An intrinsic Rashba electric field was found to form at the interface, with a corresponding parameter $\alpha = 147$ meVÅ. The Rashba spin-orbit coupling (SOC) is correlated with the interfacial charge depletion at the V/MgO interface, which is reduced (increased) in the presence of a positive (negative) electric field. However, a very large electric field of about 1 V/Å= 10 V/nm is needed to significantly affect the Rashba intrinsic field. Within the range of applied bias (voltage) present in our experiments, no major variation of $\alpha$ induced by the bias of the magnetic tunnel junction is expected. In a first-order approximation, we can consider a constant SOC for all biases used in magneto-transport experiments and just take into account the electronic structure effects on conductance, within a rigid band model. Within this picture, SOC mixes the surface $\Delta_1$ and bulk state $\Delta_2$ orbital symmetries in vanadium[2], allowing electronic transport in the V/MgO/Fe system, as shown in Fig. 1a in the main text.

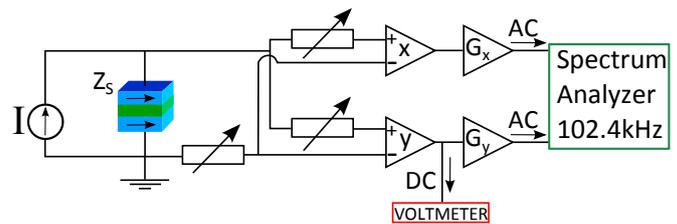

FIG. S2: Electronic set-up diagram for the shot noise and transport experiments. $Z_S$ represents the sample impedance, x and y the first amplification stage, and $G_x$ and $G_y$ are the commercial amplifiers.

## II. EXTENDED EXPERIMENTAL DETAILS AND CHARACTERIZATION

### A. Shot noise set-up and background noise removal

The schematic electronic set-up of the noise measurement system is shown in Fig. S2. This system was also tested at room temperature by measuring the thermal noise of different resistors, and using the fluctuation-dissipation theorem to calculate the Boltzmann constant, $k_B$, from a linear fit of the voltage noise as a function of resistance, $S_V = 4Tk_BR$. The results of this calibration procedure are shown in Fig. S3.

The Fano factor, $F$, is defined as

$$F = S_V^{\mathrm{exp}}/S_V, \tag{S1}$$

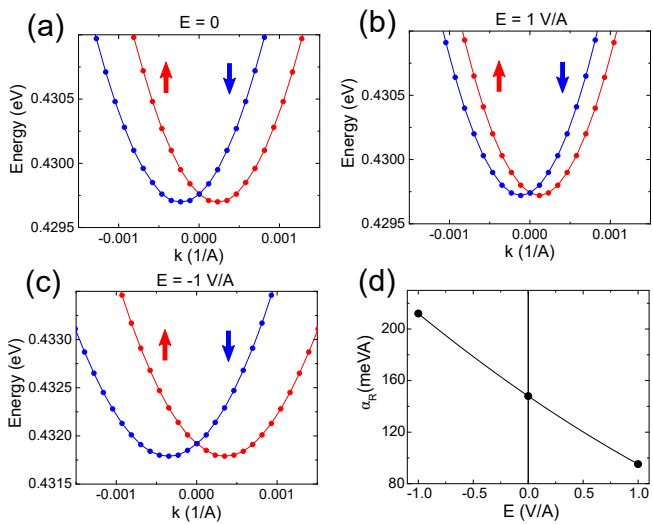

FIG. S1: Calculated band splitting due to Rashba SOC at the V/MgO interface, (a) in the absence of electric field, (b) with a positive field of $E = 1$ V/Å, and (c) with a negative one of the same intensity. The arrows indicate which band corresponds to the majority (up) and minority (down) spin populations. (d) Rashba SOC strength $\alpha$ as a function of the electric field.

Typeset by REVTEX

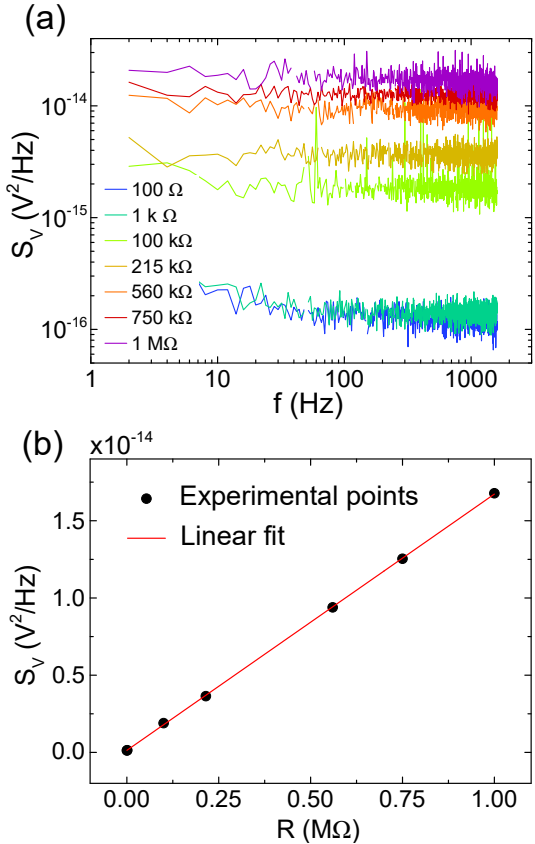

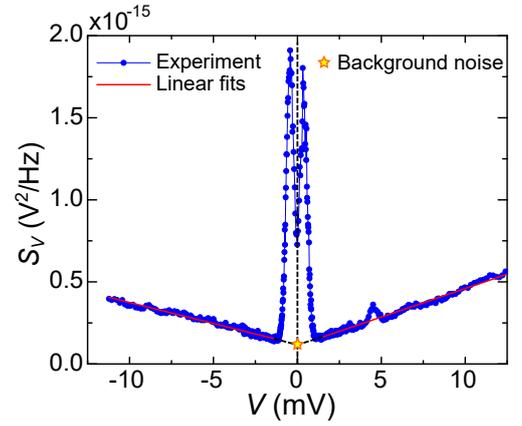

FIG. S4: Scheme of the background noise subtraction. The red lines are the linear fits performed for bias $V > \Delta$. The dashed lines extrapolate those fits to $V = 0$, where the yellow star marks the background noise level.

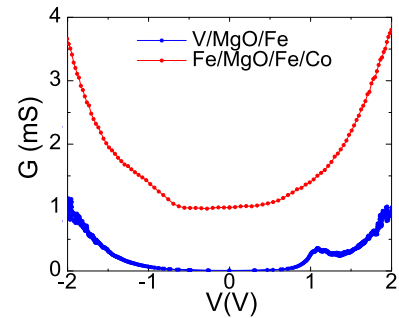

FIG. S3: (a) Power spectral density of the thermal noise for different resistors at room temperature. (b) The fit of this thermal noise as a function of resistance to obtain the Boltzmann constant, $k_B$, from the fluctuation-dissipation theorem. The slope of the linear fit in the figure is $1.66 \times 10^{-20}$, which yields $k_B = 1.38 \times 10^{-23}$ J/K, within 1% of the actual value.

FIG. S5: Conductance curves for two different samples: a V(40 nm)/MgO(2 nm)/Fe(10 nm) (S/F) and a Fe(40 nm)/MgO(2 nm)/Fe(10 nm)/Co(20 nm) (F/F), of the same lateral size ($20 \times 20$ $\mu$m$^2$), up to bias $V = 2$ V.

where $S_V^{\mathrm{exp}}$ is the voltage shot noise measured experimentally, and $S_V$ is the Poissonian voltage shot noise together with the thermal noise, given by[3]

$$\begin{aligned} S_V(V) &= S_I (dV/dI)^2 \\ &= \frac{2eI}{G^2(V)}\coth\left(eV/2k_BT\right), \end{aligned} \quad \text{(S2)}$$

where $S_I = 2eI$ is the Poissonian current noise, $-e$ is the electron charge, and $G(V)$ the differential conductance.

In Fig. 2b (in the main text), a comparison of the experimental shot noise and the expected Poissonian noise (green line, $F = 1$) is shown. That green curve uses Eq. (S2) and an additional background noise base level that comes from the electronic circuit measuring the sample signal, and cannot be removed with the cross-correlation method. This background noise is $T$ and $V$ independent, as can be seen from the $V$ dependence of the shot noise, which is in good agreement with Eq. (S2) in Fig. 2b. It is obtained from a linear fit of the experimental noise data above the effective superconducting gap, $\Delta$ (where the $V$ dependence is linear), and extrapolating the fit to $V = 0$ to find the expected zero-bias thermal noise. This is done independently for the positive and negative bias branches, and the background thermal noise is taken as the average of the crossing of the two lines with $V = 0$. This process is sketched in Fig. S4.

### B. Control experiments with different junction types

The giant shot noise is a robust phenomenon, observed in multiple (more than 10) samples, both in superconductor/ferromagnet (S/F) and S/F/F junctions. Figure S5 compares the low-temperature $G(V)$ in V/MgO/Fe S/F junction with control junctions where V is substituted by Fe to demonstrate the symmetry-induced bottleneck at the V/MgO interface shown in Fig. 1a. While $G(V = 0)$ of Fe/MgO/Fe/Co control magnetic tunnel junctions is relatively high (in the mS range for $20 \times 20$ $\mu$m$^2$ junctions[4]) and $G(V = 2$ V$)$ is three times higher, in V/MgO/Fe junctions $G(V = 0)$ is much lower, but the corresponding $G(V = 2$ V$)$ is more than 30 times higher.

Figure S6 compares the shot noise and $G$ in the sin-gle barrier S/F and F/S/F junctions of the same area ($20 \times 20$ $\mu m^2$). For the F/S/F junction, the noise signal has two sources in series, located at the Fe/MgO/V and V/MgO/Fe interfaces, which are the two symmetry filtering bottlenecks for $G$ (Fig. S6a), and the subgap excess noise is suppressed (Fig. S6b). These samples also show suppressed $G(V, \Delta)$ (Fig. S6c) and are an order of magnitude more resistive than in the S/F/F or S/F junctions for the same area due to $G$ bottleneck at the second V/MgO/Fe barrier, as discussed in the main text.

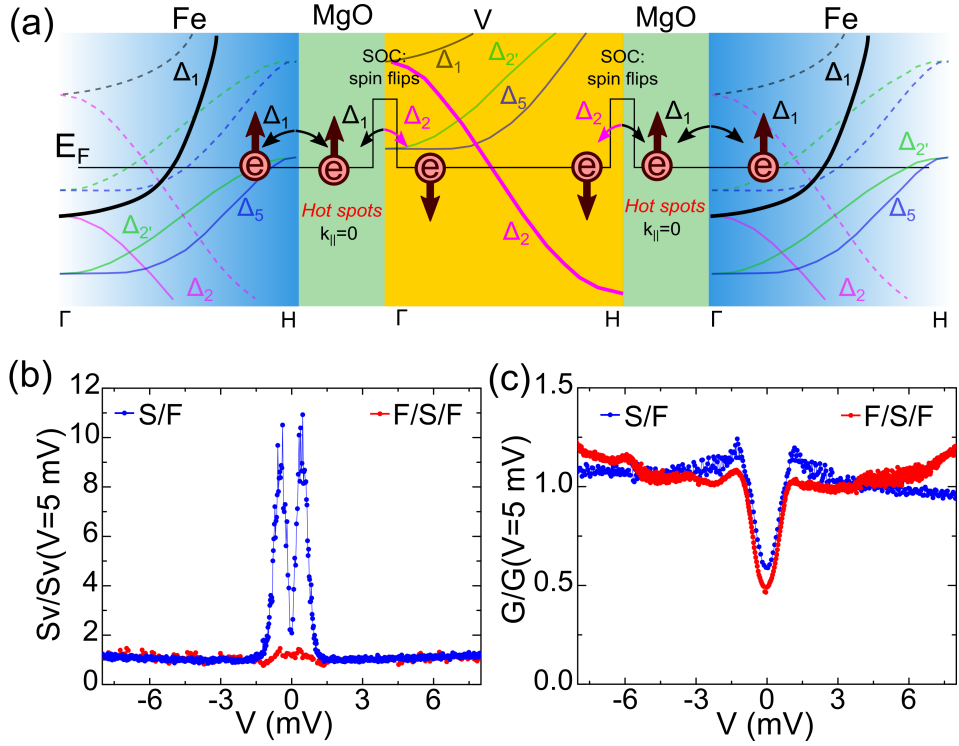

FIG. S6: (a) Sketch of the transport process in a F/S/F (Fe/MgO/V/MgO/Fe/Co) sample, showing the two bottlenecks for electrons at the MgO/V and V/MgO interfaces, where they have to undergo SOC scattering to change their orbital symmetry to be able to tunnel through the MgO. (b) Bias-dependnet shot noise in S/F (V/MgO/Fe) and F/S/F samples of the same area ($20 \times 20$ $\mu m^2$), showing how the excess subgap shot noise is suppressed in the latter. (c) Normalized $G(V)$ for two samples of the same type and area, showing the superconducting gap.

We have also carried out control measurements on V(40 nm)/MgO(2 nm)/V(40 nm) (S/S) epitaxial junctions, where the ferromagnetic Fe electrode has been substituted by superconducting vanadium. The measured shot noise and $G$ are shown in Fig. S7. The panel (a) explains why the V/MgO/V junctions do not show any signs of the Josephson effect, multiple Andreev reflections, nor excess shot noise. In the panel (e) the noise of one of these samples is compared to that of a Fe/MgO/V sample, showing no increase in the subgap shot noise, which could be expected in this "textbook example" of S/S junction. As discussed in the main text, electrons at the Fermi level of the vanadium have $\Delta_2$ symmetry. However, these cannot be transmitted through the MgO(100) barrier, as it completely filters them out[5]. The only way for quasiparticles to be transmitted between the two superconducting electrodes is by the mixing of $\Delta_2$ and $\Delta_1$ symmetries[2], which occurs due to the interfacial Rashba SOC at the two (V/MgO an MgO/V) interfaces (see Sec. I and Fig. S7a). The presence of these two SOC-controlled electron symmetry bottlenecks is evidenced by the very low normal-state $G$ in V/MgO/V junctions, close to the one of Fe/MgO/V/MgO/Fe/Co which also contain two V/MgO interfaces (see Fig. S7b, and Fig. 1b). This strong suppression of quasiparticle transmission prevents the possibility of having any Josephson current at $V = 0$, multiple Andreev reflections, and the related giant excess subgap shot noise.

In Figs. S7b and S7d we can also compare the absolute value of conductance of the control V/MgO/Au junction (where, as explained in the main text in Fig. 2f, the orbital symmetry filtering through MgO is suppressed) with the normal-state conductance of all-epitaxial junctions where orbital symmetry filtering provides a nearly exponential decrease of the normal-state conductance with the number of V/MgO interfaces (Fig. S7b).

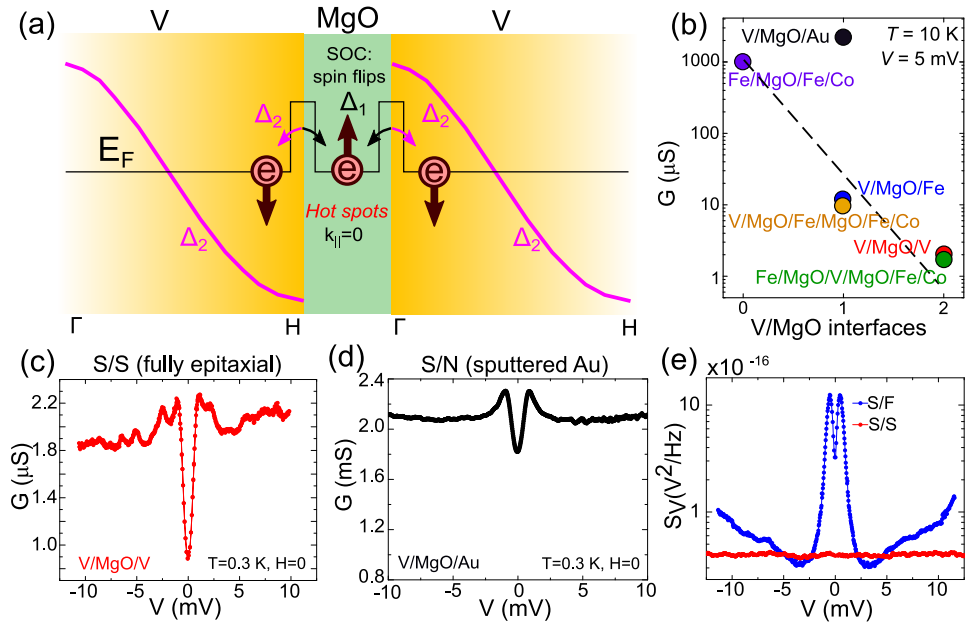

FIG. S7: (a) Sketch of the transport process in a S/S (V/MgO/V) sample, showing the two bottlenecks for electrons at the V/MgO and MgO/V interfaces, where they have to undergo SOC scattering to change their orbital symmetry to be able to tunnel through the MgO. (b) Expanded Fig. 1b of the main text showing normal-state conductance of different tunnel junctions of lateral size $20 \times 20$ $\mu m^2$ as a function of their number of V/MgO barriers. The V/MgO/Au control junction is now included, showing that it does not follow the trend of close to exponential decrease of the conductance with increasing number of V/MgO barriers (indicated by the black dashed line). (c),(d) $V$ dependence of $G$ of the epitaxially grown S/S and the nonepitaxial S/N control junction, respectively. (e) Comparison of the bias-dependent shot noise measured in a S/S and a S/F (Fe/MgO/V) junction with the same lateral sizes and MgO barrier thickness.

## III. DEPENDENCE OF THE $1/f$ NOISE ON THE MAGNETIZATION ORIENTATION AND EXTERNAL MAGNETIC FIELD

This section discusses the possible contribution of low-frequency, $f$, noise generated by superconducting vortices, induced by the external magnetic field, $H$, and stray fields created by magnetic textures in the ferromagnetic layer. We also discuss the irrelevance of this frequency-dependent noise contribution observed below a few Hz to the shot noise signals analyzed here.

First, the shot noise is generally studied in a frequency range (typically above 10 Hz) where it is nearly frequency independent (see Fig. 2c). As shown in Fig. S8, the subgap excess shot noise decreases approaching $T_C$. This $f$ independent and monotonic $T$ dependence is not the one expected for vortex-related noise, which should have a $1/f$ noise power dependence and to increase in magnitude when approaching $T_C$ due to the enhancement of vortex mobility[6]. Similarly, the monotonic reduction of the subgap noise with an increasing applied out-of-plane (OOP) field, potentially inducing vortices (see Fig. 3b), is not consistent with a scenario in which superconducting vortices are responsible for the giant shot noise.

To further investigate the possible link between the excess subgap $1/f$ noise contribution with superconducting vortices, generated by the combined effect of magnetic textures and $H$ we have studied the evolution of the normalized frequency-dependent $1/f$ noise or Hooge factor[3], $\alpha_H = S_V f A/V^2$, where A is the junction area, as a function of $V$ and in the different magnetization directions imposed by the external magnetic field. Fig-ures S9a and S9b show that near the maximum total noise, at $V = 0.44$ mV, the $f$-dependent noise contribution (below 10 Hz) has a dependence close to $1/f$, since $\alpha_H$ in nearly $f$ independent. From Figs. S9c and S9d the dependence of the $\alpha_H$ with $V$ and with the changing magnitude and the direction of **H**, shows that the measured noise is not mainly produced by the generation of superconducting vortices. It is not enhanced when the magnetization is aligned perpendicularly and the application of OOP field further diminishes it (Fig. S9d). In contrast, the behavior is more consistent with the $1/f$ noise being produced by magnetic inhomogeneities and domain wall formation. This is supported by micromagnetic simulations estimating the domain wall formation under different in-plane (IP) magnetization alignments (see Ref.[7] Supplementary Information). In the simulations, the magnetization alignment in the [100] direction created more domain walls compared to the [110] direction, which would result in a greater $1/f$ noise contribution (and therefore a higher $\alpha_H$), as observed in Fig. S9d. When $H = 1000$ Oe is applied saturating the magnetization, the domain walls disappear and the resulting $\alpha_H$ is the same for the two directions. We believe that the main reason for the excess $1/f$ noise could be these magnetic textures, effectively influencing the symmetry-dependent tunneling probability[8] and/or superconducting order parameter.

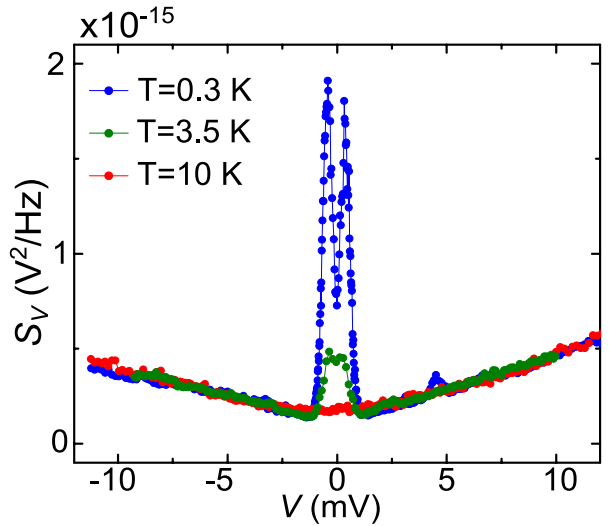

FIG. S8: Shot noise vs applied bias (measured without applied magnetic field) for a F/S junction at 0.3 K (far below $T_C$), 3.5 K (near $T_C$) and 10 K (above $T_C$). The 0.3 K and 10 K curves are shown in Fig. 2b in the main text.

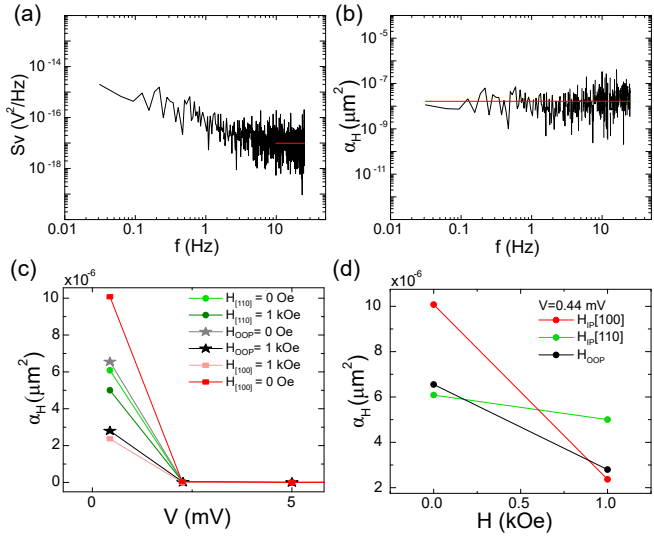

FIG. S9: (a) Typical low-frequency noise power spectrum as a function of $V$ and external magnetic field, $H$. The data are measured for a $30 \times 30$ $\mu m^2$ S/F sample at $T = 0.3$ K and $V = 0.44$ mV in (a), (b). The $1/f$ noise is clear for the lowest $f$, while above about 3 Hz the spectrum becomes flat, corresponding to the shot noise contribution. The red line is the fitted shot noise value used for the analysis. (b) Hooge factor, $\alpha_H$, calculated for each frequency point after subtracting the background shot noise. This way, we can check that the $1/f$ noise approximation is reasonable as we see that $\alpha_H$ is mostly $f$ independent in the range under study. The red line shows the fitted value used for $\alpha_H$ in our analysis. (c) $\alpha_H(V)$ for different IP (in the [110] and [100] directions) and OOP applied fields. (d) $\alpha_H(V = 0.44$ mV) as a function of $H$ for the different directions.

## IV. THEORY FOR TRANSPORT NOISE

### A. Distribution function of net charge as recursive process

In this work, we extend a widely-used approach for obtaining $I - V$ characteristics and $G$ in superconducting junctions[9,10] to compute the giant effective charge in the subgap transport. The initial approach[10] is based on ballistic transport as in the Landauer theory that the electron/hole density distribution function travels inside the normal region without dissipation. The theory does not take into account the pair-correlation distribution, without which the current fluctuations cannot be computed inside the normal region. To circumvent the problem, Dieleman et al.[11] instead evaluated the average effective charge, $|q|$, carried by the scattering states into the leads. In the tunneling regime, $|q|$ has been established to be equivalent to the Fano factor. Here, we generalize the theory that has neglected the normal reflection[11] to consider superconducting junctions with normal reflections, spin polarization, and the left-right junction asymmetry.

We use the energy convention that the energy $E$ is measured from the center of the Fermi levels of the source and drain leads. The source/drain (or left/right) lead has the Fermi level at $eV/2$ and $-eV/2$, respectively, with the bias $V$. Then the prior work[10], to describe transport in S/normal region/S junctions (in the absence of SOC and spin polarization), can be summarized using the distribution functions $f_{\rightarrow,\leftarrow}(E)$ for the left/right-moving carriers, respectively

$$f_{\rightarrow}(E) = A_L(E)[1 - f_{\leftarrow}(-E + eV)] + B_L(E)f_{\leftarrow}(E) + T_L(E)f_{\text{FD}}(E - eV/2), \quad \text{(S3)}$$

$$f_{\leftarrow}(E) = A_R(E)[1 - f_{\rightarrow}(-E - eV)] + B_R(E)f_{\rightarrow}(E) + T_R(E)f_{\text{FD}}(E + eV/2), \quad \text{(S4)}$$

where $A_\alpha(E)$, $B_\alpha(E)$, and $T_\alpha(E)$, are the scattering coefficients for the Andreev reflection, normal reflection, and transmission, respectively, from the interface $\alpha = L, R$. The initial distribution function is given in terms of the Fermi-Dirac function, $f_{\rm FD}(E) = (e^{E/k_BT} + 1)^{-1}$.

By reconstructing the solution $f_\eta(E)$ ($\eta = \rightarrow, \leftarrow$) recursively in terms of the number $N$ of reflections from either of the interfaces

$$f_\eta(E) = \sum_{N=0}^{\infty} f_\eta^N(E), \tag{S5}$$

we have the equivalent conditions as in Eqs. (S3) and (S4) with

$$f_\rightarrow^{N+1}(E) = A_L(E)[1 - f_\leftarrow^N(-E + eV)] + B_L(E) f_\leftarrow^N(E) \tag{S6}$$
$$f_\leftarrow^{N+1}(E) = A_R(E)[1 - f_\rightarrow^N(-E - eV)] + B_R(E) f_\rightarrow^N(E) \tag{S7}$$

with the initial distributions at $N = 0$ of

$$f_\rightarrow^0(E) = T_L(E) f_{\rm FD}(E - eV/2) \text{ and } f_\leftarrow^0(E) = T_R(E) f_{\rm FD}(E + eV/2). \tag{S8}$$

## B. Transport noise calculation from distribution function

In previous approaches[9,10], only the distribution of net charge is kept in $f_\eta^N(E)$. To evaluate the fluctuation of effective charges in transport, we need to decompose $f_\eta^N(E)$ into the components $g_{\nu\eta m}^N(E)$ with the moving direction $\eta$ as carrier kind $\nu = e, h$ (electron, hole), resolved with the charge transfer index $m$ (in the unit of the electron charge $-e$) carried in each transmitted partial wave.

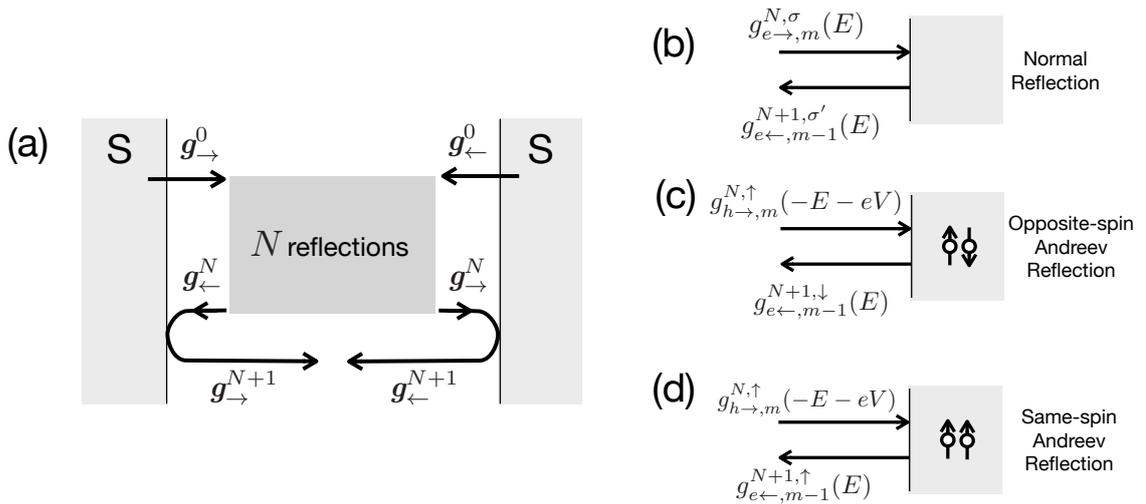

FIG. S10: (a) Schematic diagram for the recursive process for the distribution function $\boldsymbol{g}_\eta^N$ ($\eta = \rightarrow, \leftarrow$) in the S/normal region/S junction. After the $(N + 1)$-th reflection from either of the interfaces, $g_\eta^{N+1}$ is generated. (b) In a normal reflection from the right S, the distribution $g_{e\rightarrow,m}^{N,\sigma}(E)$ [right-moving ($\rightarrow$), of spin $\sigma$, carrier of electron ($e$), and the accumulated $m$ electron-charge to the current upon transmission] is reflected as the same carrier ($e$) to a spin $\sigma'$. (c) The usual opposite-spin Andreev reflection converts an incoming hole to a reflected electron. (d) SOC enables the same-spin Andreev reflection.

Figure S10 schematically shows how the electron ($\nu = e$) and hole ($\nu = h$) flux, $g_{\nu\eta m}^N$, after $(N + 1)$-th reflection generate the next order distribution. For example, in the reflection off the right interface, we show two processes that result in an electron flux afterwards ($\nu' = e$) with the normal and the Andreev reflection in (b) and (c), respectively. In (b), the right moving flux ($\eta = \rightarrow$) becomes $\eta' = \leftarrow$ and the electron flux remains ($\nu = \nu' = e$) after a normal reflection. The net charge deposited on the right reservoir is $\Delta m = -1$ since when the electron is reflected there is a lack of an electron charge as opposed to a transmitted charge. In (c), to result in an electron emission after an Andreev reflection, a hole has to be incident ($\nu = h$). In the case of the Andreev reflection, the lead receives two holes as opposed to one transmitted hole, therefore an extra hole charge to the lead and $\Delta m = -1$. In the presence of the SOC, it is straightforward to include the spin-flip scattering as illustrated in (d). We denote the scattering functions as $\boldsymbol{A}$, $\boldsymbol{B}$ and $\boldsymbol{T}$ as boldfaced for matrix in spin space. Similarly, we use the distribution function $\boldsymbol{g}$ as a vector in spin space. To summarize, we have the recursion relation,

$$\boldsymbol{g}_{e\leftarrow,m}^{N+1}(E) = \boldsymbol{A}_R(E)\,\boldsymbol{g}_{h\to,m+1}^{N}(-E-eV) + \boldsymbol{B}_R(E)\,\boldsymbol{g}_{e\to,m+1}^{N}(E). \tag{S9}$$

Similarly, for the hole emission and for the reflection from the left-interface, we have

$$\boldsymbol{g}_{h\leftarrow,m}^{N+1}(E) = \boldsymbol{A}_R(E)\,\boldsymbol{g}_{e\to,m-1}^{N}(-E-eV) + \boldsymbol{B}_R(E)\,\boldsymbol{g}_{h\to,m-1}^{N}(E) \tag{S10}$$

$$\boldsymbol{g}_{e\to,m}^{N+1}(E) = \boldsymbol{A}_L(E)\,\boldsymbol{g}_{h\leftarrow,m-1}^{N}(-E+eV) + \boldsymbol{B}_L(E)\,\boldsymbol{g}_{e\leftarrow,m-1}^{N}(E) \tag{S11}$$

$$\boldsymbol{g}_{h\to,m}^{N+1}(E) = \boldsymbol{A}_L(E)\,\boldsymbol{g}_{e\leftarrow,m+1}^{N}(-E+eV) + \boldsymbol{B}_L(E)\,\boldsymbol{g}_{h\leftarrow,m+1}^{N}(E). \tag{S12}$$

The scattering coefficients in matrices $\boldsymbol{A}$, $\boldsymbol{B}$ and $\boldsymbol{T}$ can be obtained by solving the single-interface S/F junction with interfacial Rashba SOC[12]. The results in the main text have been obtained with coefficients solved numerically. For simplicity and completeness of the theory, we present at the end of this section analytic results in a simplified limit for $\boldsymbol{A}$, $\boldsymbol{B}$, $\boldsymbol{T}$, and the resulting Fano factor.

The initial distributions are

$$\begin{aligned}
g_{e\to,+1}^{0,\sigma}(E) &= T_L(E) f_{\text{FD}}(E - eV/2 - \mu_{R\sigma}) \\
g_{e\leftarrow,-1}^{0,\sigma}(E) &= T_R(E) f_{\text{FD}}(E + eV/2 - \mu_{L\sigma}) \\
g_{h\to,-1}^{0,\sigma}(E) &= T_L(E)[1 - f_{\text{FD}}(E - eV/2 - \mu_{R\sigma})] \\
g_{h\leftarrow,+1}^{0,\sigma}(E) &= T_R(E)[1 - f_{\text{FD}}(E + eV/2 - \mu_{L\sigma})],
\end{aligned} \tag{S13}$$

with $\mu_{\alpha\sigma}$ as the spin-dependent chemical potential in each lead $\alpha$ to account for the spin polarization. All other $g_{\nu\eta m}^{0,\sigma}$ are zero. This formulation is equivalent to the previous works[9,10] with the distribution function given as

$$f_\eta^\sigma(E) = \sum_{N=0}^{\infty} \sum_{m=-\infty}^{\infty} g_{e\eta,m}^{N\sigma}(E). \tag{S14}$$

The total current carried by the electrons and the holes is

$$I = \sum_{m=-\infty}^{\infty} I_m \text{ with } I_m = \frac{m}{2eR_0} \sum_{N=0}^{\infty} \sum_{\eta=\to,\leftarrow} \sum_{\nu=e,h} \sum_{\sigma} \int T_\alpha(E) g_{\nu\eta m}^{N\sigma}(E) dE, \tag{S15}$$

with the normal resistance $R_0$. The index $m$ in $I_m$ is the accumulated charge transfer (in unit of $-e$) of all electron and hole Cooper pairs and the eventual transmission of an electron or a hole into either of the reservoirs with the transmission $T_\alpha(E)$. With the translational symmetry parallel to the interface the theory is quantized with a transverse wave vector, and the formulation is easily extended to dimensions greater than one. One needs to define $\boldsymbol{A}$, $\boldsymbol{B}$ and $\boldsymbol{T}$ for each transverse wave vector, repeat the above recursion processes separately, and sum over the transverse wave vector in observables such as Eqs. (S14) and (S15). We define the effective magnitude of the charge transfer $\langle|q|\rangle$ as weighted according to the current as

$$\langle|q|\rangle = \frac{\sum_{m=-\infty}^{\infty} |m| I_m}{\sum_{m=-\infty}^{\infty} I_m}. \tag{S16}$$

We use $\langle|q|\rangle$ to interpret the Fano factor as discussed in the text.

### C. Absence of giant Fano factor without proximitized superconductivity

As argued in the main text, this formulation makes it particularly transparent as to why a proximitized superconductivity should be present for a giant Fano factor. With its absence, the Andreev reflection is active on only one interface, and the successive Andreev reflections alternate between the electron/hole pairs on the interface. Then, the resulting $|m|$ cannot exceed 2 due to the charge cancellation, and the Fano factor remains conventional.

We demonstrate the fact from the structure of the above recursive relations. We set the left lead as non-superconducting, $\boldsymbol{A}_L(E) = 0$. From Eq. (S10), a right-moving electron distribution $\boldsymbol{g}_{e\to,m}^{N}$ is Andreev-reflected to a left-moving hole distribution $g_{h\leftarrow,m+1}^{N+1}$, which is then Andreev-reflected via Eq. (S9) back to the right-moving electron distribution $g_{e\rightarrow,m}^{N+2}$ without any net change in $m$. With the initial distributions given for $m = \pm 1$ in Eq. (S13), $g_{\nu\eta m}^N = 0$ for $|m| > 2$ and the effective charge $\langle |q| \rangle$, Eq. (S16), remains conventional. The same can be argued for the hole distribution. Figure 4**c** in the main text shows the Fano factor bounded by 2 when the proximity-induced superconducting gap in the F region, $\Delta_F$, is set to zero (blue curve).

### D. Effective 1-dimensional model for scattering coefficients and Fano factor

Finally, we demonstrate the procedure in the quasi-1D limit of our system of an S/normal region/S junction. We first solve the Hamiltonian of an normal region/S junction given as

$$\hat{H} = \begin{pmatrix} \hat{H}_e & \Delta_S \Theta(z) I_{2\times 2} \\ \Delta_S^* \Theta(z) I_{2\times 2} & \hat{H}_h \end{pmatrix}, \tag{S17}$$

where the single-particle Hamiltonian for electrons is

$$\hat{H}_e = -\frac{\hbar^2}{2m^*}\frac{\partial^2}{\partial z^2} - \mu - \frac{\Delta_{xc}}{2}\hat{\mathbf{m}}\cdot\hat{\boldsymbol{\sigma}}\Theta(-z) + (V_0 d + \alpha k_y \hat{\sigma}_x)\delta(z), \tag{S18}$$

and for holes

$$\hat{H}_h = -\hat{\sigma}_y \hat{H}_e^* \hat{\sigma}_y. \tag{S19}$$

Here, the propagation direction is along the z-axis and $k_y$ is a transverse wave vector treated as a constant. $m^*$ is the effective mass and $\hat{\boldsymbol{\sigma}}$ are Pauli matrices. $\Delta_S$ is the spin-singlet superconducting gap for vanadium, $\Theta(z)$ is the Heaviside step function and $\delta(z)$ is the Dirac delta function. $\Delta_{xc}$ is the exchange spin splitting and $\hat{\mathbf{m}}$ denotes the magnetization orientation. The interface is characterized by an effective barrier height $V_0$ over the thickness $d$ and the Rashba SOC with the strength $\alpha$. It is convenient to introduce the dimensionless parameters for potential barrier and SOC strength,

$$Z = \frac{V_0 d m^*}{\hbar^2 k_F}, \ Z_F = \frac{\alpha k_y m^*}{\hbar^2 k_F}. \tag{S20}$$

For a transparent demonstration of the procedure, we consider a simplified limit with the Andreev approximation[13] and with zero spin polarization ($\Delta_{xc} = 0$), where tractable analytic solutions are available. By matching the boundary condition across the N/S junction, we obtain the solutions for the scattering coefficients explicitly as: $A_{\uparrow\uparrow} = A_{\downarrow\downarrow} = |a_\sigma|^2$, $A_{\uparrow\downarrow} = A_{\downarrow\uparrow} = |a_{\overline{\sigma}}|^2$, $B_{\uparrow\uparrow} = B_{\downarrow\downarrow} = |b_\sigma|^2$, $B_{\uparrow\downarrow} = B_{\downarrow\uparrow} = |b_{\overline{\sigma}}|^2$, $C_{\uparrow\uparrow} = C_{\downarrow\downarrow} = |c_\sigma|^2 \beta \Theta(|E| - \Delta_S)$, $C_{\uparrow\downarrow} = C_{\downarrow\uparrow} = |c_{\overline{\sigma}}|^2 \beta \Theta(|E| - \Delta_S)$, $D_{\uparrow\uparrow} = D_{\downarrow\downarrow} = |d_\sigma|^2 \beta \Theta(|E| - \Delta_S)$, $D_{\uparrow\downarrow} = D_{\downarrow\uparrow} = |d_{\overline{\sigma}}|^2 \beta \Theta(|E| - \Delta_S)$ with $\beta = \sqrt{E^2 - \Delta_S^2}/E$, and $\boldsymbol{T} = \boldsymbol{C} + \boldsymbol{D}$, where

$$\begin{aligned}
a_\sigma &= \sqrt{1-\beta^2}\left[1 + \beta + 2\beta(Z^2 + Z_F^2)\right]/\Omega, \\
a_{\overline{\sigma}} &= -4\beta Z_F Z\sqrt{1-\beta^2}/\Omega, \\
b_\sigma &= -2\beta\left[(1+\beta)(Z^2 + Z_F^2 + iZ) + 2\beta(Z^2 - Z_F^2)(Z^2 - Z_F^2 + iZ)\right]/\Omega, \\
b_{\overline{\sigma}} &= 2i\beta Z_F\left[(1+\beta)(2iZ-1) + 2\beta(Z^2 - Z_F^2)\right]/\Omega, \\
c_\sigma &= \sqrt{2(1+\beta)}\left[(1+\beta)(1-iZ) + 2\beta(Z^2 + Z_F^2 - iZ(Z^2 - Z_F^2))\right]/\Omega, \\
c_{\overline{\sigma}} &= -iZ_F\sqrt{2(1+\beta)}\left[1 + \beta - 2\beta(Z^2 - Z_F^2 + 2iZ)\right]/\Omega, \\
d_\sigma &= iZ\sqrt{2(1-\beta)}\left[1 + \beta + 2\beta(Z^2 - Z_F^2)\right]/\Omega, \\
d_{\overline{\sigma}} &= iZ_F\sqrt{2(1-\beta)}\left[1 + \beta - 2\beta(Z^2 - Z_F^2)\right]/\Omega
\end{aligned} \tag{S21}$$

with $\Omega = [1 + \beta + 2\beta(Z + Z_F)^2][1 + \beta + 2\beta(Z - Z_F)^2]$. Here, the energy $E = 0$ is defined at the Fermi energy of the S/normal region junction. To convert the above expressions to a biased S/normal region/S junction we shift the reference energy to the middle of the two Fermi energies, *i.e.* $\boldsymbol{A}_{L/R}(E) = \boldsymbol{A}(E \mp eV/2)$, etc. in a symmetric junction. The energy-dependent coefficients for typical barrier parameters are plotted in Fig. S11a.

Using these explicit coefficients for the normal region/S junction, we can compute distribution functions $g_{\nu\eta m}^{N\sigma}(E)$ in the S/normal region/S junction by following the above recursive procedures. The charge current and the Fano factor are then obtained from Eqs. (S14)-(S16). The resulting bias-dependent Fano factor is shown in Fig. S11b.

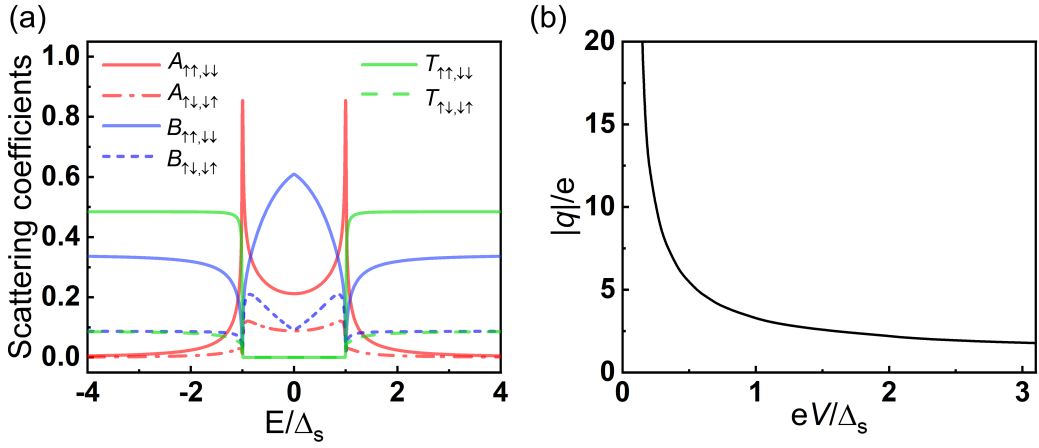

FIG. S11: (a) Scattering coefficients as a function of energy in the 1D normal region/S junction with the interfacial potential barrier $Z = 1$ and the Rashba SOC $Z_F = 0.6$. (b) Fano factor or, equivalently, the effective charge ratio $|q|/e$, as a function of bias in the 1D S/normal region/S junction with the same barrier parameters as given in (a), while $T = 0.075\ T_c$, and $\Delta_S = 1.76\ k_B T_c$ for a vanadium sample.

### E. Dependence of Fano factor on the barrier potential

In Fig. S12, we present an additional comparison of the Fano factor curves from Fig. 4c for different values of the barrier potential $Z$. Although the barrier parameters significantly influence the conductance profile, their effect on the Fano factor remains minimal. We conclude that the Fano factor is primarily determined by the value of the proximity-induced gap $\Delta_F$, at least in the range of $Z$ parameters used to fit the conductance (see Fig. S13).

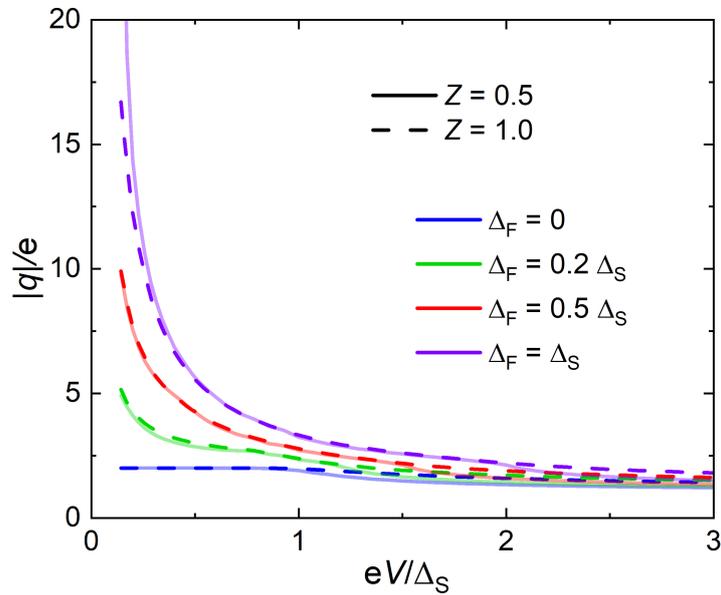

FIG. S12: Calculated Fano factor or, equivalently, the effective charge ratio $|q|/e$, as a function of applied bias for different barrier strength $Z$ (solid and dashed lines) and proximity-induced gap $\Delta_F$.

## F. Conductance fitting and parameter selection

The experimental $G-V$ curve is fitted using the theoretical model described above in Fig. S13. The proximity-induced gap is estimated by $\Delta_F/\Delta_S = \Delta_{\text{Fe}}/\Delta_{\text{Au}} - 1 \approx 20\%$, where $\Delta_{\text{Fe}}$ and $\Delta_{\text{Au}}$ are the measured gap of F/I/S and N/I/S junctions, respectively, in Fig. 2d. We apply a broadening function to the calculated conductance curve to account for inelastic scattering, which smooths out the secondary peaks. It can be seen in Fig. S13 that the choice of $Z=1, P=0.7$ provides the best fit, showing excellent agreement with the experimental curve.

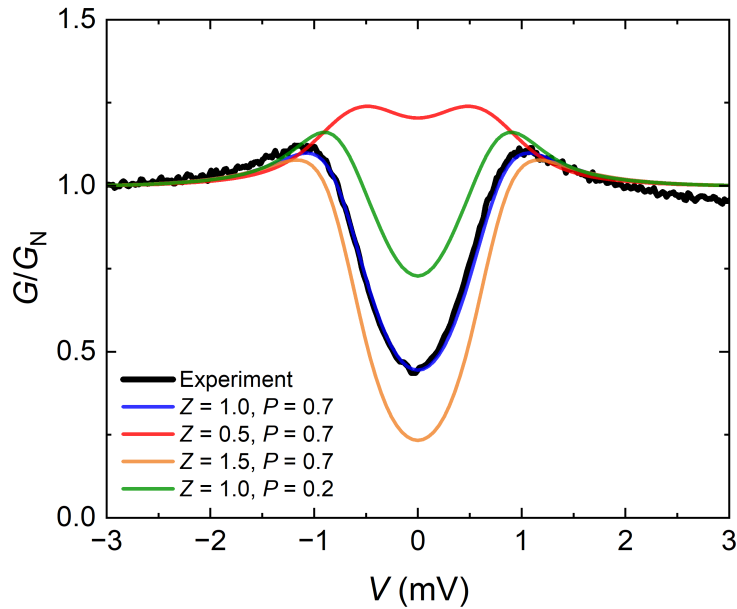

FIG. S13: Bias-dependent conductance: experimental measurements and theoretical fits with varying parameters. The proximity-induced gap $\Delta_F = 0.2\Delta_S$, and the Rashba SOC $\lambda = 1.2$.